\documentclass[ preprint,
 superscriptaddress,
 amsmath,amssymb,
 aps,
]{revtex4-2}
\usepackage{graphicx}
\usepackage{physics}
\usepackage{xspace}

\usepackage[encapsulated]{CJK}
\usepackage{hyperref}
\usepackage{color}

\usepackage[normalem]{ulem}

\makeatletter
\newsavebox{\@brx}
\newcommand{\llangle}[1][]{\savebox{\@brx}{\(\m@th{#1\langle}\)}%
  \mathopen{\copy\@brx\kern-0.5\wd\@brx\usebox{\@brx}}}
\newcommand{\rrangle}[1][]{\savebox{\@brx}{\(\m@th{#1\rangle}\)}%
  \mathclose{\copy\@brx\kern-0.5\wd\@brx\usebox{\@brx}}}
\makeatother

\usepackage{forloop,ifthen} 

\usepackage{xifthen}
\newcommand{\refAppendix}[6]{#1
  \ifthenelse{\isempty{#2}}%
    {}
    {\protect\cite{#2}}
    #3\protect\ref{#4}#5#6\xspace
}

\def\rhosq{{\mbox{$\hat{\rho}_\text{S}$}}}

\begin{document}

\title{Machine Learning for Quantum State Tomography: Robust Covariance Matrix 
Estimation for Squeezed Vacuum States with Thermal Noise}

\author{Juan Camilo Rodríguez,$^1$ Hsien-Yi Hsieh,$^3$ Hua-Li Chen,$^{1}$ Ole Steuernagel,$^{2}$ Chien-Ming Wu,$^{2}$ and Ray-Kuang Lee$^{1,2,3,4*}$ }
\affiliation{$^1$ Department of Physics, National Tsing Hua University, Hsinchu 30013, Taiwan\\
$^2$ Institute of Photonics Technologies, National Tsing Hua University, Hsinchu 30013, Taiwan\\
$^3$ Center for Theory and Computation, National Tsing Hua University, Hsinchu 30013, Taiwan\\
$^4$ Center for Quantum Science and Technology, Hsinchu 30013, Taiwan}

\email{*rklee@ee.nthu.edu.tw} 

\begin{abstract} 
We present a supervised machine learning-based method using convolutional neural networks to estimate the covariance matrix of Gaussian quantum states in the presence of thermal noise. Unlike computationally intensive density matrix reconstructions, our machine learning-based method allows for the reconstruction of impure squeezed vacuum states using sparse measurements of quadrature sequences based on a model employing a two-component state mixed together from thermal and squeezed thermal states. The method achieves high fidelity and precision, notably also at high squeezing levels, while offering an effective characterization of physical quantities and accurately estimating the covariance matrix. We benchmark our machine against experimental data of single-mode squeezed vacuum states, demonstrating its accuracy and capability to quantify experimental degradation to squeezing and purity. We experimentally verify that our covariance matrix estimation exhibits robustness to state degradation induced by thermal state admixtures. We provide a method for lightweight, compact, and complete representation of lab-generated Gaussian states and lay the foundation for extending real-time quantum state tomography for thermal multi-component Gaussian states to multi-mode systems.
\end{abstract}
\maketitle

\section{Introduction}
Quantum information science has rapidly emerged as a result of advances in the control and
manipulation of quantum systems, leading to groundbreaking applications in fields such as quantum
cryptography \cite{BENNETT20147,Ekert1991,Gisin2002}, quantum sensing \cite{Girolami2014}, quantum
communications \cite{Kimble2008}, and quantum simulation \cite{Georgescu2014}. These applications
aim to leverage quantum mechanics for secure data transmission, ultra-precise measurements, and the
simulation of complex quantum systems that are computationally intractable with classical
methods. The development of quantum processors, capable of performing operations significantly
faster than classical machines, has established a new paradigm in information processing
\cite{supremacy}, driving progress in both quantum computation and information protocols. As quantum
technology advances, developing efficient methods to extract information encoded in complex quantum
states is essential for achieving precision in quantum measurements and enhancing the reliability of
quantum information applications.

Quantum state tomography (QST) aims to reconstruct the information of a quantum state created under
controlled experimental conditions \cite{d2003quantum,Lvovsky2009}. Traditionally, maximum
likelihood estimation has been used to identify the most likely quantum state that best fits the
experimental data \cite{Hradil,Badurek2004}. However, maximum likelihood estimation is
computationally intensive, particularly for high-dimensional systems, where the computational cost
grows exponentially with the Hilbert space dimension. This limitation has motivated alternative
approaches to achieve faster, more efficient and reliable tomography. Machine learning algorithms
have emerged as new candidates to perform tomography, enhancing speed, efficiency and accuracy. These algorithms range from generative machines \cite{Carrasquilla2019,Lloyd2018,Nathan2017} over
deep and residual neural networks \cite{Hsieh2022,Hsieh2022(2),Hsien2023,Hsien2024} to the latest advances in artificial intelligence that are being used to enhance the power of quantum state
reconstruction.

In this paper, we present a machine learning-based QST method with a particular focus on
reconstructing squeezed vacuum states of light, allowing us to analyze systems which
traditionally require the handling of large Hilbert spaces bases. These states have applications
ranging from quantum information protocols to enhancing the sensitivity of interferometric
gravitational wave detectors \cite{Gravwaves, FDS}. We adopt the covariance matrix to represent
 the states reducing computational cost while capturing the essential properties of such Gaussian
states. We have reported machine learning-based single-mode state reconstruction via reconstruction
of the density matrix, $\hat{\rho}$, to achieve a complete characterization of squeezed vacuum
states~\cite{Hsieh2022}. Our method reported here advances this process by reconstructing the
covariance matrix through fast measurements, enabling real-time QST~\cite{Hsieh2022} while maintaining computational efficiency for practical applications. Compared with our earlier density–matrix CNN approach~\cite{Hsieh2022}, the present covariance-matrix formulation is substantially lighter and easier to deploy, while preserving reconstruction accuracy (see Sec.~\ref{sec:data_analysis}).

Our approach is validated using experimental data from squeezed vacuum states produced in the lab, benchmarking results against traditional methods. Our approach promises to be applicable to multi-mode Gaussian states reconstructions, where the covariance matrix framework offers significant advantages in handling multi-mode Hilbert spaces, facilitating the extraction of entanglement information in real time. By advancing these techniques, we move closer to enabling real-time quantum control, with implications for quantum computation and information processing
\cite{Nielsen_Chuang_2010}.

This paper is organized as follows. Section~2 reviews the covariance-matrix formalism for Gaussian states, introduces the two-component thermal-mixture model used throughout, and presents the parametrization that enforces the Heisenberg constraint. Section~3 details the supervised pipeline—synthetic data generation, training-set construction, and the ResNet-CNN architecture—and discusses hardware feasibility. Section~4 describes the experimental squeezed vacuum state reconstruction, including the SQ–ASQ degradation curve and purity trends. Section~5 presents our analysis workflow: pre-processing of quadrature records, covariance-matrix-based inference and a comparison with density-matrix reconstructions. Section~6 concludes and outlines future directions.

\section{Quantum Covariance Matrix}

In reconstructing quantum states, it is beneficial to identify a sparse representation that
efficiently captures the properties of the state while preserving the constraints imposed by quantum
mechanics. Single-mode states have traditionally been reconstructed using the density matrix
$\hat{\rho}$, but this approach requires a large number of parameters and computational resources,
especially when dealing with high-dimensional Hilbert spaces \cite{Hsieh2022}. For Gaussian states a
description using their covariance matrix, $\boldsymbol{\sigma}$, is an excellent alternative,
offering a compact yet complete representation of the quantum state. Shifting to the covariance
matrix significantly reduces computational complexity. By comparison, using the density matrix
requires truncation in Fock space to make the problem tractable~\cite{Ferraro_2005, Weedbrook2012, Adesso_2014,
  serafini2017quantum}.

The elements of the covariance matrix, $\boldsymbol{\sigma}$, are
\begin{equation}\label{eq:sigmaDef}
\sigma_{ij}=\langle\{\Delta \hat R_i,\Delta \hat R_j\}\rangle,\quad \Delta \hat R_i=\hat R_i-\langle \hat R_i\rangle
\end{equation}
where $\mathbf{R}= (\hat{x}_1,\hat{p}_1,\hat{x}_2,\hat{p}_2, ...)^\mathrm{T}$ is a vector of
canonical operators $(\hat{x}_j, \hat{p}_j)$ for $N$ continuous variable modes,~$j$; the indices
$i, j = 1, \dots, 2N$ enumerate the dimensions of the associated $2N$-dimensional phase
space. $\boldsymbol{\sigma}$ is a real, symmetric, and positive definite matrix that quantifies the
correlations between pairs of canonical operators~\cite{Ferraro_2005, Weedbrook2012, Adesso_2014,
  serafini2017quantum}.

Unlike a Hilbert space-based density matrix, the covariance matrix defines the quantum state with
reference to phase space, allowing us to take advantage of its symplectic structure, particularly of
symplectic transformations which preserve the canonical commutation
relations~\cite{Ferraro_2005}. For our purposes, the most important symplectic transformations are
rotations and squeezing transformations which are elements of the symplectic group
Sp$(2N,\mathbb{R})$ that provides us with mathematically simple operations and their implementation
for the manipulation of covariance matrices~\cite{serafini2017quantum}.

Quantum physics imposes a constraint for positive semi-definiteness 
on the covariance 
matrix~$\boldsymbol{\sigma}$ of expression~(\ref{eq:sigmaDef}) when it is added to the 
symplectic form. This constraint is denoted by the symbol `$\geq 0$' and has the explicit form 
\begin{equation}\label{Eq 2} 
\boldsymbol{\sigma}+ \text{i} \boldsymbol{\Omega} \geq 0
\text{, \quad with the symplectic form: }\quad
\boldsymbol{\Omega} =  \begin{pmatrix}
\mathbf{0} & \mathbf{I} \\
- \mathbf{I} & \mathbf{0} 
\end{pmatrix} 
 \; .
\end{equation}
This condition ensures that $\boldsymbol{\sigma}+i\boldsymbol{\Omega}$ is positive semidefinite, which is the covariance matrix formulation of the \textit{Heisenberg uncertainty principle} \cite{Weedbrook2012, Adesso_2014, serafini2017quantum, Ferraro_2005}. For the single-mode case, Eq.~(\ref{Eq 2}) is equivalent to requiring $\det\boldsymbol{\sigma} \geq 1$, the familiar determinant form of the uncertainty relation. In the multi-mode case, the full matrix inequality must be enforced, as it guarantees that all symplectic eigenvalues satisfy $\nu_j \geq 1$. The machine developed in this work is explicitly designed to respect these constraints during the training process and in applications.

For a single-mode squeezed state, the covariance matrix is a $2\times2$ real symmetric matrix that
stores the variances and covariances of the quadrature operators $\hat{x}$ and $\hat{p}$. For a
given phase angle $\theta$, these quadratures are defined as
\begin{equation} \label{eq:quadratures} \hat{x}(\theta)=\dfrac{1}{\sqrt2}\left(\hat{a}^\dag
 \text{e}^{ \text{i} \theta}+\hat{a}  \text{e}^{- \text{i} \theta}\right),\hspace{0.1cm}\hat{p}(\theta)
=\dfrac{1}{\sqrt2 \text{i}}\left(\hat{a}^\dag  \text{e}^{ \text{i} \theta}-\hat{a}  \text{e}^{- \text{i} \theta}\right) \;,
\end{equation}
and they correspond to real and imaginary parts of a quantized single electromagnetic field-mode
with respect to the local oscillator reference phase
$\theta$~\cite{Scully_Zubairy_1997,serafini2017quantum,Banaszek1997}. Statistical measurements of
these quadratures provide sufficient information to fully reconstruct a quantum
state~\cite{Lvovsky2009}.  We refer to every set of measurements of the
quadratures~(\ref{eq:quadratures}) for phases~$\theta \in [0, \pi]$ as a \textit{quadrature
  sequence}~\cite{Hsieh2022}.

The covariance matrix $\boldsymbol{\sigma}$ of a Gaussian state can be diagonalized through local
(symplectic) rotations to align with suitable alignment angle $\theta_0$~\cite{Williamson1936}. For a
single-mode system with alignment angle~$\theta_0$ this removes the off-diagonal coherences between
$x$ and $p$ yielding the explicit form
\begin{equation}
\label{eq:DiagForm}
\boldsymbol{\sigma}_{\text{diag}}(\theta_0) =
\boldsymbol{S}^\mathrm{T} \boldsymbol{\sigma} \boldsymbol{S} 
=
\begin{bmatrix}
    \sigma_{xx} & 0 \\
    0 & \sigma_{pp}
\end{bmatrix}
=
\begin{bmatrix}
    \displaystyle\min_{\theta}2{\langle (\Delta \hat{x}(\theta))^2 \rangle} & 0 \\
    0 & \displaystyle\max_{\theta}2{\langle (\Delta \hat{p}(\theta))^2 \rangle}
\end{bmatrix}
=
\begin{bmatrix}
    \text{e}^{-2r} & 0 \\
    0 & \text{e}^{2r}
\end{bmatrix},
\end{equation}
where $r$ is the squeezing parameter and $\langle (\Delta \hat{x}(\theta))^2 \rangle$, $\langle (\Delta
\hat{p}(\theta))^2 \rangle$ are the quadrature variances in $x$ and $p$, respectively. 

This diagonalization simplifies the representation through covariance matrices even further, making computations extremely efficient. In this setting, the alignment angle $\theta_0$ does not only diagonalize $\boldsymbol{\sigma}$; it also coincides with the squeezing angle, $\theta_0=\phi/2$, easily computed by minimizing the variance, thereby specifying the phase point of minimal variance, $\langle(\Delta\hat x(\theta_0))^2\rangle$, and its conjugate counterpart of maximal variance, $\langle(\Delta\hat p(\theta_0))^2\rangle$.

Even though a full density matrix reconstruction provides more complete information, including the
full characterization of degradation processes~\cite{Hsieh2022}, in the context of Gaussian-state
applications, working with the covariance matrix is more practical and straightforward in describing
the required observables. It captures the essential second-order statistics and offers a significant
computational advantage, as it reduces the number of parameters and allows for a smaller, more
efficient convolutional neural network. Our method is particularly well suited for Gaussian
information processing, and its compact form makes it easy to integrate as an in-line toolbox for
fast and reliable characterization of the key features of Gaussian states, especially in cases
involving complex noise conditions~\cite{Kobayashi}.

A Gaussian state can be fully described by its covariance matrix in the diagonal
form~(\ref{eq:DiagForm}) and its alignment angle~$\theta_0$~\cite{Williamson1936}. For squeezed
vacuum states this is the minimum number of parameters that allows us to fully specify a state in
this framework. When dealing with two-component Gaussian distributions, such as~(\ref{Eq 6}) below, our
approach ensures that the second-order estimation remains accurate.  For more general (mildly non-Gaussian)
states, the determination of covariance matrices allows us to filter out the Gaussian part of the
state; this simplification does not give a complete reconstruction of the state but can be useful in
some circumstances.

\subsection{Two-component Gaussian states}

The processing of noisy measurement data is essential in realistic implementations of quantum state
tomography.  Given their success in supervised learning tasks for signal processing, convolutional
neural networks (CNNs) have shown strong potential for `de-noising' experimental measurements and
extracting features from noisy data accurately.

The description of thermal noise has previously been included~\cite{Hsieh2022} as a single-component
Gaussian state,~$ \rhosq$, formed from effective thermal states, $\hat \rho_{\text{th}}(n)$,
where $n$ is a fit-parameter which behaves like an effective photon number, squeezed by the
single-mode squeezing operator
$\hat{S}(r,\phi)=\exp[\frac{1}{2}(r \text{e}^{- \text{i} \phi}\hat{a}^2 - r \text{e}^{ \text{i}
  \phi}\hat{a}^{\dag2})]$
\begin{equation}\label{Eq 5}
  \rhosq(r,n,\phi)=\hat{S}^{\dag}(r,\phi)\hat{\rho}_{\text{th}}(n)\hat{S}(r,\phi).
\end{equation}
This formulation encompasses different limiting cases: for $r=0$ it reduces to a thermal state,
for $n=0$ it yields a pure squeezed vacuum, and for $r \neq 0$ and $n \neq 0$ it describes a general
squeezed thermal state.
Compared to our previous work~\cite{Hsieh2022}, this work demonstrates an alternative 
machine learning-based method by using the two-component state~$\hat{\rho}_{\text{noisy}}$ of the form
\begin{equation}\label{Eq 6} \hat{\rho}_{\text{noisy}}(r,n,\phi,\epsilon) =
(1-\epsilon)\rhosq(r,n,\phi) + \epsilon \hat{\rho}_{\text{th}}(n) \; ,
\end{equation} with a sliding noise weight $\epsilon \in [0,0.5]$. Note that state $
\hat{\rho}_{\text{noisy}}$ 
is not strictly Gaussian since it is the sum of two different Gaussian component states; also note, the
same value for $n$ is used in both of its components
$\rhosq(r,n,\phi)$ and $\hat{\rho}_{\text{th}}(n)$.

\subsection{Theory on extracting the covariance matrix}
The estimation of~$\boldsymbol{\sigma}$ must obey the
\textit{Heisenberg uncertainty} constraint in Eq.~(\ref{Eq 2}) for quantum covariance matrices. This
\textit{uncertainty} constraint poses a stricter requirement than positive definiteness
of~$\boldsymbol{\sigma}$ by itself. To enforce it, we introduce the auxiliary diagonal matrix
$\boldsymbol{A} = \text{diag}\left(-\frac{\sigma_{xx} + 1}{\sigma_{pp} + 1}, 1\right)$. When added
to~$\boldsymbol{\sigma}$, $\boldsymbol{A}$~is designed to generate the strictly positive-definite
matrix, $\boldsymbol{\tau} > 0$, where
\begin{equation}\label{Eq 7} \boldsymbol{\tau} = \boldsymbol{\sigma} + \boldsymbol{A} \; .
\end{equation}

With this construction, we factorize $\boldsymbol{\tau}$ via the Cholesky decomposition as $\boldsymbol{\tau}=\boldsymbol{L}\boldsymbol{L}^{\!\top}$, where $\boldsymbol{L}$ is a lower triangular Cholesky matrix. Rather than predicting $\boldsymbol{\sigma}$ directly, the network outputs the entries of $\boldsymbol{L}$. We then reconstruct $\boldsymbol{\tau}$ and invert the transform to obtain $\boldsymbol{\sigma}=\boldsymbol{\tau}-\boldsymbol{A}$, which by construction satisfies $\boldsymbol{\sigma}+i\boldsymbol{\Omega}\ge0$. This guarantees that the estimated covariance matrix obeys the \textit{Heisenberg uncertainty principle} and is physically valid for quantum state characterization.

\begin{figure}[t] \centering \includegraphics[width= 11.0cm]{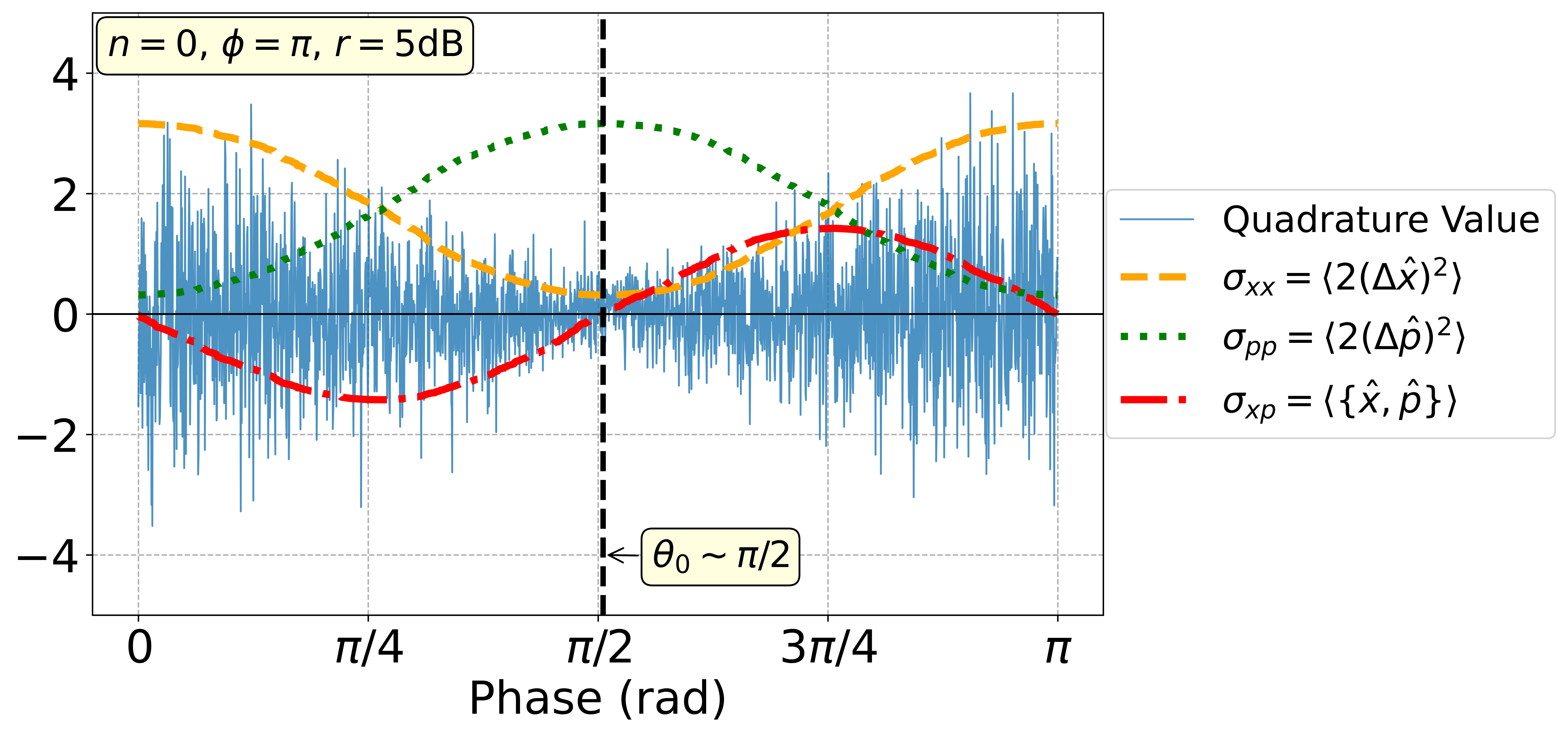}
    \caption{Quadrature sequences (solid curves) for a pure squeezed–thermal state [Eq.~(\ref{Eq 5})]. The corresponding covariance matrix elements $\sigma_{xx}(\theta)$, $\sigma_{pp}(\theta)$ and $\sigma_{xp}(\theta)$ [Eq.(~\ref{eq:sigmaDef})], are overlaid as dashed lines. The vertical dashed line marks the alignment angle $\theta_0=\pi/2$, where the covariance matrix takes a diagonal form.}
    \label{Fig1}
\end{figure}

\section{Methodology and Machine Architecture}

In our machine learning framework, the target of the network is the covariance matrix in its diagonal form, as shown in Eq.~(\ref{eq:DiagForm}). Fig.~\ref{Fig1} illustrates how a quadrature sequence from the squeezed–thermal model $\rhosq$ of Eq.~(\ref{Eq 5}) relates to the covariance–matrix elements; $\sigma_{xx}(\theta)$, $\sigma_{pp}(\theta)$, and $\sigma_{xp}(\theta)$. These curves are overlaid, and the vertical dashed line marks the alignment angle $\theta_0=\pi/2$, where $\boldsymbol{\sigma}$ becomes diagonal.

Quadrature sequences are subsequently generated by sampling the values of the rotated phase space quadrature operator ${\hat{x}(\theta)}$ using the \textit{NumPy} library function \texttt{random.normal()}~\cite{numpy_normal}. They are then fed to a neural network that maps them, through a series of layers, to a physically valid covariance matrix using the parameterization described in Sec.~2.2. This generates a training dataset consisting of sequences of tuples $\{\{\hat{x}(\theta)\}, \theta \}$, with 2048 points sampled for each state. Consequently, our inputs are of size $2048 \times 2$ per state for the phase values $\theta \in [0, \pi]$. Parameters are sampled as $n \in [0, 1]$ for thermal photon numbers, $\phi \in [0,\pi]$ and squeezing levels $r \in [0, 15]$~dB.

To populate the dataset, we use the two-component noisy Gaussian model of Eq.~(\ref{Eq 6}), $\hat{\rho}_{\text{noisy}}(r,n,\phi,\epsilon)$, which provides a mixture of a squeezed-thermal and a thermal state sharing the same $n$. The quadrature sequences used for training, 
using the \textit{QuTiP 4.7.5} library, are sampled from this distribution, ensuring that the network is trained 
directly on data reflecting the experimental presence of thermal noise. Because the 
\textit{QuTiP 4.7.5} library represents states in a truncated Hilbert space, we impose a cut-off chosen such that the total state probability remains above $0.9999$, ensuring numerical accuracy.

\newpage

\begin{figure}[t] \centering \includegraphics[width=12.0cm]{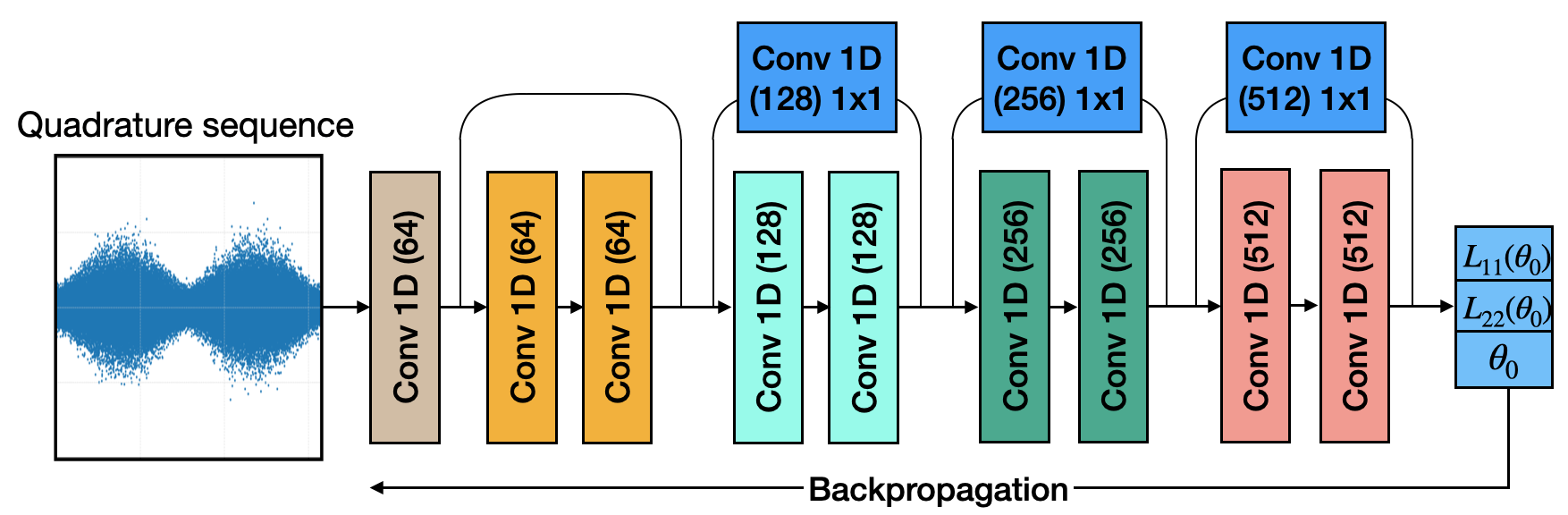}
  \caption{Architecture for single-mode QST reconstructing the covariance matrix in diagonal
form~(\ref{eq:DiagForm}). The architecture consists of multiple residual blocks with batch normalization and ReLU activations. The kernel size is 7 for every layer except for the small connections which have $1\times1$ filters. The number of filters is enclosed in parentheses. Stride is always `2'. Once a triangular matrix $\boldsymbol{L}$ is obtained as output, the transformed covariance matrix $\boldsymbol{\tau}=\boldsymbol{L}\boldsymbol{L}^\dag$ is generated and applied to train the weights of the network via the backpropagation loop.}
    \label{Fig2}
\end{figure}
\textit{Machine Learning architecture}.---Fig.~\ref{Fig2} illustrates the full architectural layout of the machine.  We use the TensorFlow \cite{tensorflow2015-whitepaper} framework choosing the ResNet-CNN architecture~\cite{he2015deepresiduallearningimage} to construct our machine. This choice mitigates the vanishing gradient problem, allowing for deeper layers that can utilize symmetries in the quadrature sequences.

The input to the network is a matrix of size $2048 \times 2$, where the first column contains the quadrature values, and the second column holds the associated phase values randomly sampled from the interval $\theta \in [0, \pi]$. This design connects quadrature values with their corresponding phases, enabling the machine to extract phase-dependent patterns from data.

The machine processes inputs through a series of $2$D convolutional layers. Each convolutional block includes a convolution layer followed by a ReLU activation function which introduces non-linearity to the output. This structure is repeated for multiple layers, with subsequent layers connected via skip connections, see Fig.~\ref{Fig2}, (as per ResNet-CNN architecture) to ensure efficient gradient flow in the backpropagation step. The final layer estimates the Cholesky matrix $\boldsymbol{L}$ in diagonal form, along with the alignment angle $\theta_0$. The loss function is evaluated by using the mean squared error between the resulting estimated and reference covariance matrices. Backpropagation updates the weights via gradient descent, and convergence is achieved when the loss reaches a value of 6.6 $\times 10^{-3}$.

\textit{Hardware feasibility}.---Relative to the density–matrix CNN approach of Ref.~\cite{Hsieh2022}, which required predicting a truncated Hilbert–space representation, our present machine reconstructs only the covariance matrix. This change makes the network substantially lighter: the number of trainable parameters is reduced by more than a factor of four (439{,}875 vs.\ 1{,}786{,}795), and the total memory footprint decreases from $\approx 6.8$\,MB to $\approx 1.7$\,MB. Beyond being more compact, this architecture is naturally suited to embedded real-time implementations, since the full model fits entirely in the on-chip memory of modern FPGA devices and can be executed as a streaming pipeline \cite{Wu2025}. Thus, compared to the previous density-matrix CNN, our method not only preserves accuracy but also enhances the feasibility of real-time continuous-variable QST.

\section{Experimental Squeezed Vacuum State Reconstruction}

The generation of a single-mode squeezed vacuum state is achieved using a bow-tie optical parametric oscillator (OPO) with a periodically poled potassium titanyl phosphate (PPKTP) crystal, as commonly implemented in continuous-variable quantum optics experiments \cite{Lvovsky2009,Hsieh2022}. Squeezed light from the OPO is superposed with a local oscillator (LO) field via a beam splitter in a balanced homodyne detection setup. By scanning the phase of the LO for $\theta \in [0,\pi]$, we collect quadrature sequences for a full state characterization. The noise profile of the vacuum (ground) state serves as a reference to calibrate our measurements.

\textit{Degradation curve}.—We characterize performance using the degradation curve, which tracks squeezing versus anti-squeezing, both in dB. By controlling the noise weight parameter $\epsilon$ in Eq.~(\ref{Eq 6}), we train with datasets for varying levels of two-component Gaussian noise. We test different models on these datasets with $\epsilon =\{0, 0.01, 0.02, $ $0.03, 0.04, 0.05\}$. To identify the best-performing model, we compute the mean squared error (MSE) between the predicted squeezing and anti-squeezing levels $(\mathrm{SQ}, \mathrm{ASQ})$ and the corresponding experimental measurements, this minimum is reached at $\epsilon=0.01$ (see Table \ref{table}), and was used to perform all estimations. This shows that we can account for part of the total noise as some mixture with thermal states as in the two-component states of Eq.~(\ref{Eq 6}).
\begin{table} \centering
    \begin{tabular}{|c|c|c|c|c|c|c|} \hline $\epsilon$ & 0.00 & 0.01 & 0.02 & 0.03 & 0.04 & 0.05 \\
\hline MSE & 0.94 & 0.49 & 1.17 & 6.53 & 4.64 & 3.91 \\ \hline
    \end{tabular}
    \caption{Mean squared error (MSE) between model and experiment for varying noise levels
($\epsilon$).}
    \label{table}
\end{table}

\begin{figure}[t] \centering \includegraphics[width=8.0cm]{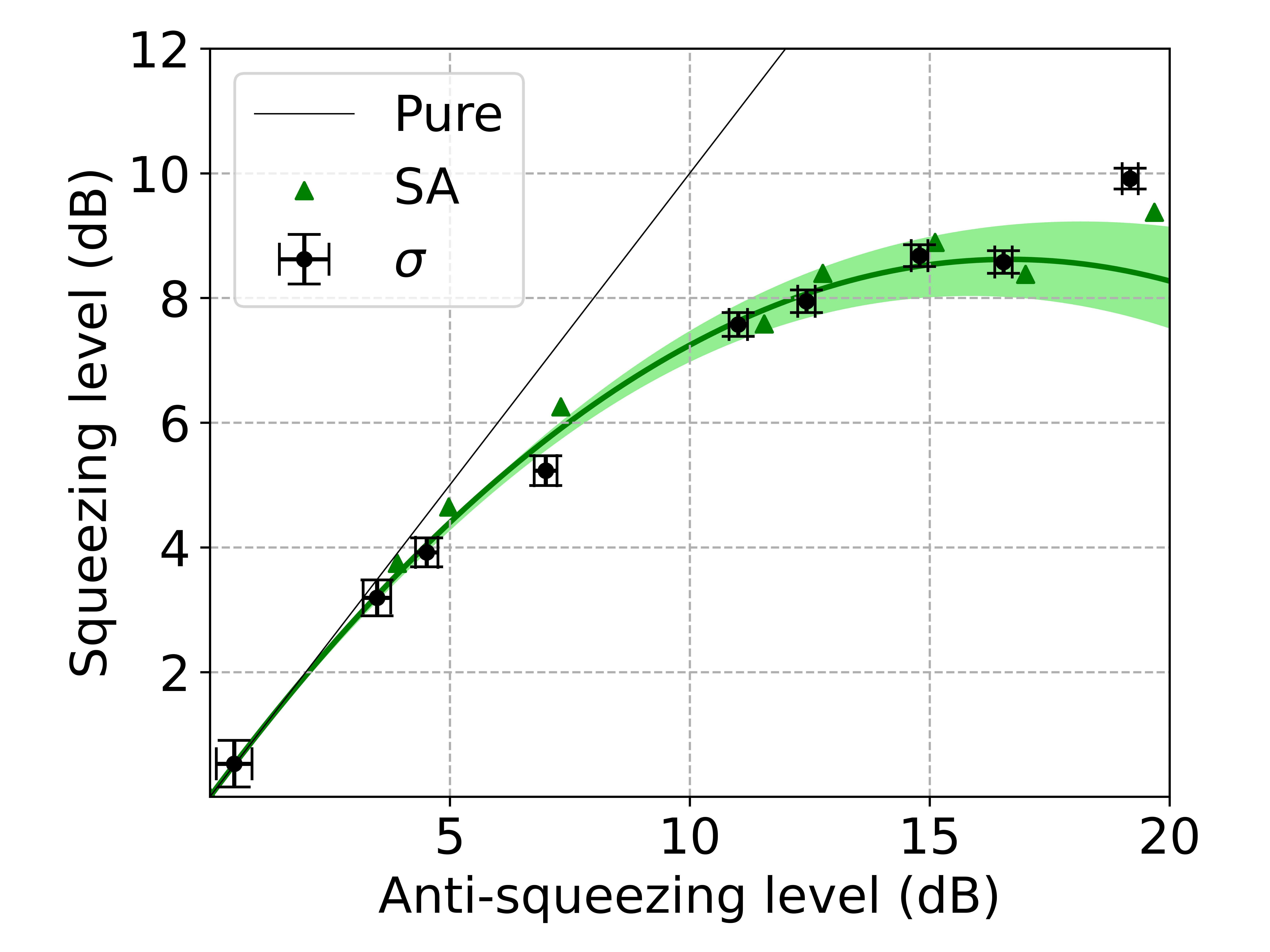}
  \caption{Degradation of single-mode squeezed vacuum states as a function of anti-squeezing. The thin gray line shows the ideal pure case without degradation. The green solid curve is a fit to data obtained from a power spectrum analyzer (SA), and the surrounding shaded green area indicates $\pm1$ standard deviation of that fit. Green dots mark the experimental values extracted from the SA measurements. The covariance-matrix-based reconstructions (`$\sigma$') are shown with black error bars indicating $\pm1$ standard deviation. The close agreement between our covariance matrix estimations and the fit to SA data demonstrates that our method accurately captures degradation arising from losses and phase noise, even when, for large anti-squeezing, we enter
  the domain of two-component states of Eq.~(\ref{Eq 6}).}
  \label{Fig3}
\end{figure}

The results of our covariance matrix-based approach, denoted by `$\sigma$',
are presented in Fig.~\ref{Fig3}, in which losses induced by the environment, 
imperfections in the laser field, and thermal vacuum noise are reflected in the 
observed degradation, which increases with the anti-squeezing level. The thin gray line 
represents an ideal squeezed vacuum state, green points represent experimental 
estimates of the squeezing and anti-squeezing levels, with the shaded regions 
indicating \mbox{$\pm1$ standard} deviations.

\textit{Purity}.---For Gaussian states, the purity is
\(p \equiv [\det\boldsymbol{\sigma}]^{-1/2}\), with \(\boldsymbol{\sigma}\) the covariance matrix.
It equals \(p=1\) only for pure states and decreases below 1 as the state becomes mixed. The noise
processes relevant here—such as thermal admixtures considered in Eq.~(\ref{Eq 6})—monotonically increase
\(\det\boldsymbol{\sigma}\) and therefore reduce \(p\) \cite{Weedbrook2012,Adesso_2014,serafini2017quantum}.
Accordingly, the purity \(p\) serves as a compact scalar indicator of degradation in our experiments.

\begin{figure}[t] \centering \includegraphics[width=8.0cm]{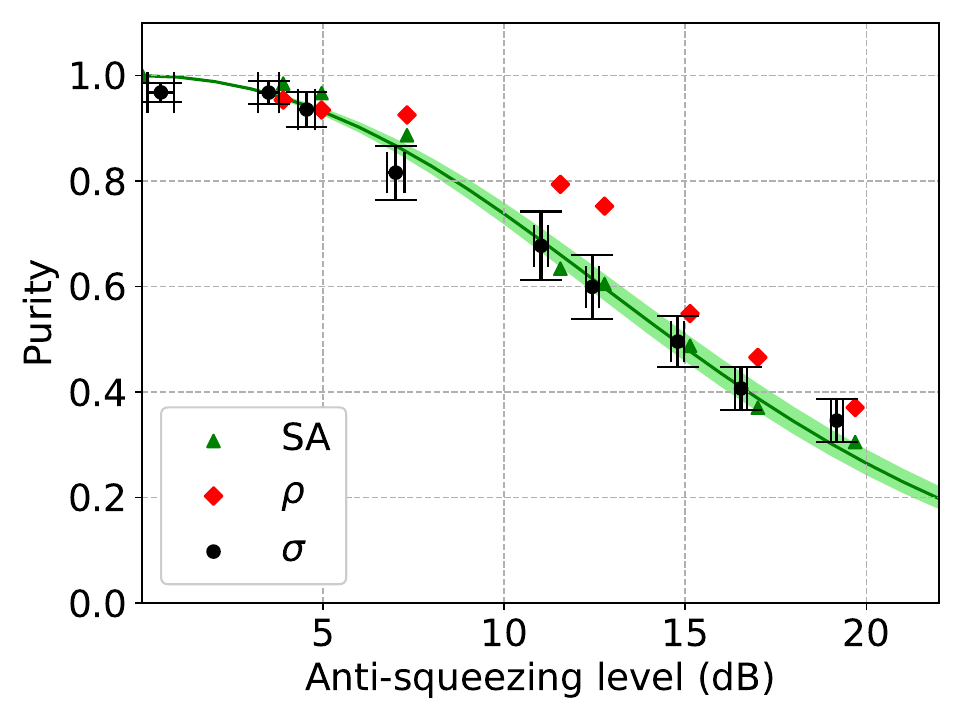}
  \caption{Purity of single-mode squeezed states as a function of anti-squeezing. The green curve is a fit to data obtained from a power spectrum analyzer (SA), with the shaded green area indicating  $\pm1$ standard deviation of the fit. Green markers represent the experimental data points extracted from SA measurements. Black vertical error bars (covariance-matrix-based estimates, `$\sigma$') denote  $\pm1$ standard deviation, while the red markers (density-matrix reconstructions, `$\rho$',) are shown for comparison. Higher anti-squeezing levels correspond to stronger degradation and hence reduced purity.}
    \label{Fig4}
\end{figure}

Fig.~\ref{Fig4} shows the relationship between purity and anti-squeezing. As expected, higher anti-squeezing levels correspond to stronger degradation, reflected in reduced purity, consistent with the trend observed in the degradation curve of Fig.~\ref{Fig3}. Error bars represent \(\pm1\) standard deviation over quadrature sub-samples. For comparison, we also include density-matrix estimations (denoted in $\rho$) from Ref.~\cite{Hsieh2022} together with the experimental fit; both are compatible with the covariance-matrix ($\sigma$) results (including error bars), though the density-matrix approach tends to overestimate purity. Importantly, the purity obtained from the covariance matrix method is not only accurate but also faster to compute than when reconstructing the density matrix, making it especially suitable for efficient single-scan tomography even at anti-squeezing levels approaching \(\sim20\,\mathrm{dB}\). In Sec. 5.3 Fidelity is analyzed as an additional degradation indicator.

\section{Data Analysis}
\label{sec:data_analysis}

\subsection{Pre-processing}

A single experimental run of raw homodyne data consists of pairs 
$\{(x_k,\theta_k)\}_{k=1}^N$ with $N \sim 3\times10^6$, where $x_k$ is the quadrature value and 
$\theta_k \in [0,\pi]$ exploits the $\pi$-rotation symmetry of squeezed states. 
From this dataset, we uniformly select a quadrature sequence of $N=2048$ elements, 
which is sufficient to perform reliable reconstructions (see Ref.~\cite{Hsieh2022}). 
The variance of the vacuum state is used for calibration, allowing all observables to be reported 
on an absolute dB scale. This procedure step quantifies the spread of our estimates within a single 
experimental dataset and thus characterizes the precision of the method.

To account for residual imperfections in the quadrature statistics, we model the data with the 
two-component mixture $\hat\rho_{\mathrm{noisy}}(r,n,\phi,\epsilon)$ of Eq.~(\ref{Eq 6}), a combination of a 
squeezed-thermal state and a thermal state (sharing the same $n$), which offers 
a description of the presence of thermal noise consistent with the quadrature distribution 
while enforcing the quantum constraint via the transformation in Eq.~(\ref{Eq 7}).

\subsection{Covariance-matrix-based inference} The quadrature data are fed to a \mbox{ResNet-CNN} that outputs the elements of a Cholesky factor $L$; the covariance matrix is recovered inverting Eq.~(\ref{Eq 7}) $\boldsymbol\sigma = \boldsymbol L\boldsymbol L^{\!\top} - \boldsymbol A$ to guarantee $\boldsymbol{\sigma}+i\boldsymbol\Omega\ge0$ ($\det \boldsymbol\sigma\geq1$). From
$\boldsymbol{\sigma}$ we extract
\begin{equation} \text{SQ} = 10 \log_{10} \left( \boldsymbol{\sigma}_\text{diag}(\theta_0)_{xx}
\right), \quad \text{ASQ} = 10 \log_{10} \left( \boldsymbol{\sigma}_\text{diag}(\theta_0)_{pp}
\right)
    \label{eq:derived}
\end{equation}
where $\boldsymbol{\sigma}_\text{diag}(\theta_0)_{xx}$ and
$\boldsymbol{\sigma}_\text{diag}(\theta_0)_{pp}$ are the diagonal elements of the covariance matrix in Eq.~(\ref{eq:DiagForm}).

The \emph{degradation curve} in Fig.~\ref{Fig3} plots the squeezing level (SQ) against the
anti-squeezing level (ASQ). For ideal states, we have the thin gray line along
$\text{SQ}=\text{ASQ}$. Because the experiment naturally interacts with many noise sources (e.g.,
optical loss, phase jitter, electronic noise), increasing the squeezing level leads to greater
degradation; this behavior has been experimentally measured up to $\sim22\,\mathrm{dB}$. We include
the optimal fitting curve obtained by the orthogonal distance regression \cite{Hsieh2022} together
with the corresponding standard deviation in the shaded area.

To create statistically independent replicates we randomly draw $2048$ phase points uniformly from $[0,\pi]$ and retain the corresponding quadratures; this uniform down-sampling is repeated with different random seeds until 1000 \text{sub-sample datasets} are gathered. For experimental data we 
calculate their standard deviation and display vertical error bars indicating $\pm1$ standard deviation. 
The small size of these error bars, together with the close agreement to the experimental points, 
demonstrates high precision and stability even in the presence of strong decoherence. The size of the deviation is comparable to that inferred from the power spectrum analyzer data. Estimating the covariance matrix is advantageous since the parameters that we aim to estimate to reconstruct the state 
are directly obtained from the covariance matrix elements of Eqs. (8). To
determine squeezing and anti-squeezing levels as well as purity, the quantum state reconstruction 
based on covariance matrices requires fewer transformations to fully reconstruct the 
two-component state $\hat\rho_{noisy}$. This analysis confirms excellent agreement 
with experimental data even in the presence of strong decoherence; 
our standard deviation assessment further corroborates the precision and reliability of our estimates.

\subsection{Further degradation information—Fidelity} 

We have shown that our machine reconstructs experimental data well; additionally, we also assess how closely the reconstructed states approximate the actual ones. To quantify robustness, we evaluate the fidelity on $6000$ independently generated two-component states defined by Eq.~(\ref{Eq 6}).

In continuous-variable quantum systems, the fidelity between two quantum states is a standard metric to evaluate the quality of state reconstruction quantifying the closeness between two quantum states. In the specific case of two single-mode Gaussian states $\rhosq(\boldsymbol{\sigma})$ and $\hat{\rho}_0(\boldsymbol{\sigma_0})$ (with zero mean), we use the fidelity expression \cite{Adesso_2014}

\begin{equation} F(\hat\rho_S, \hat\rho_0) = \frac{2}{\sqrt{\Delta + \delta} - \sqrt{\delta}} ,
\label{eq:fidelity_gaussian}
\end{equation} where
\[ \Delta := \det(\boldsymbol{\sigma} + \boldsymbol{\sigma}_0), \quad \delta := (\det
\boldsymbol{\sigma} - 1)(\det \boldsymbol{\sigma}_0 - 1),
\] $\hat\rho_S$ is the simulated density matrix and $\hat\rho_0$ is the reconstructed one. The
fidelity takes values in the range $0 \leq F(\hat\rho_S, \hat\rho_0) \leq 1$, with $F = 1$ indicating
perfect reconstruction, for \textit{pure} and \textit{mixed} states.

\begin{figure}[t]
\centering
\includegraphics[height=5cm]{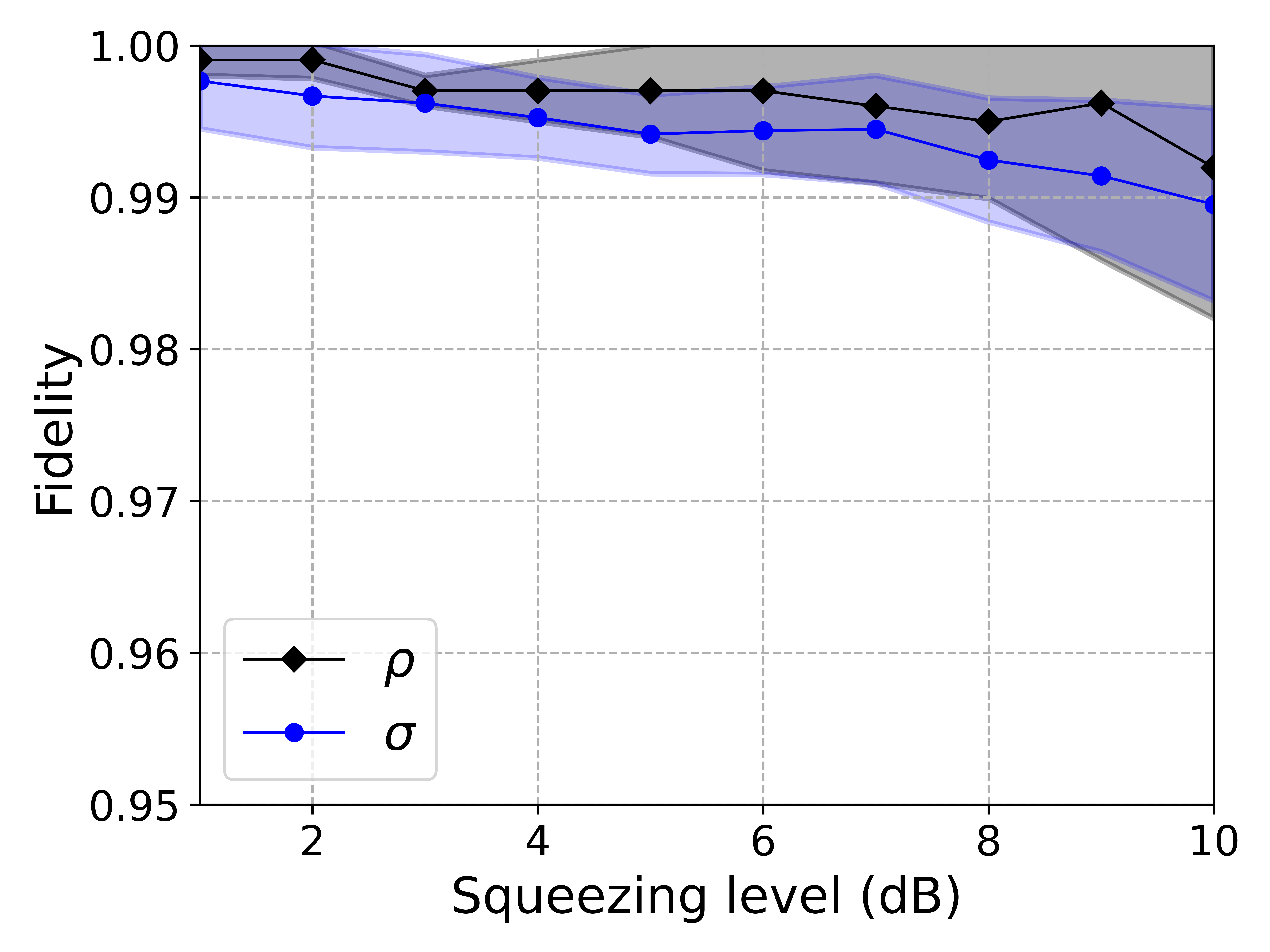}
\includegraphics[height=5cm]{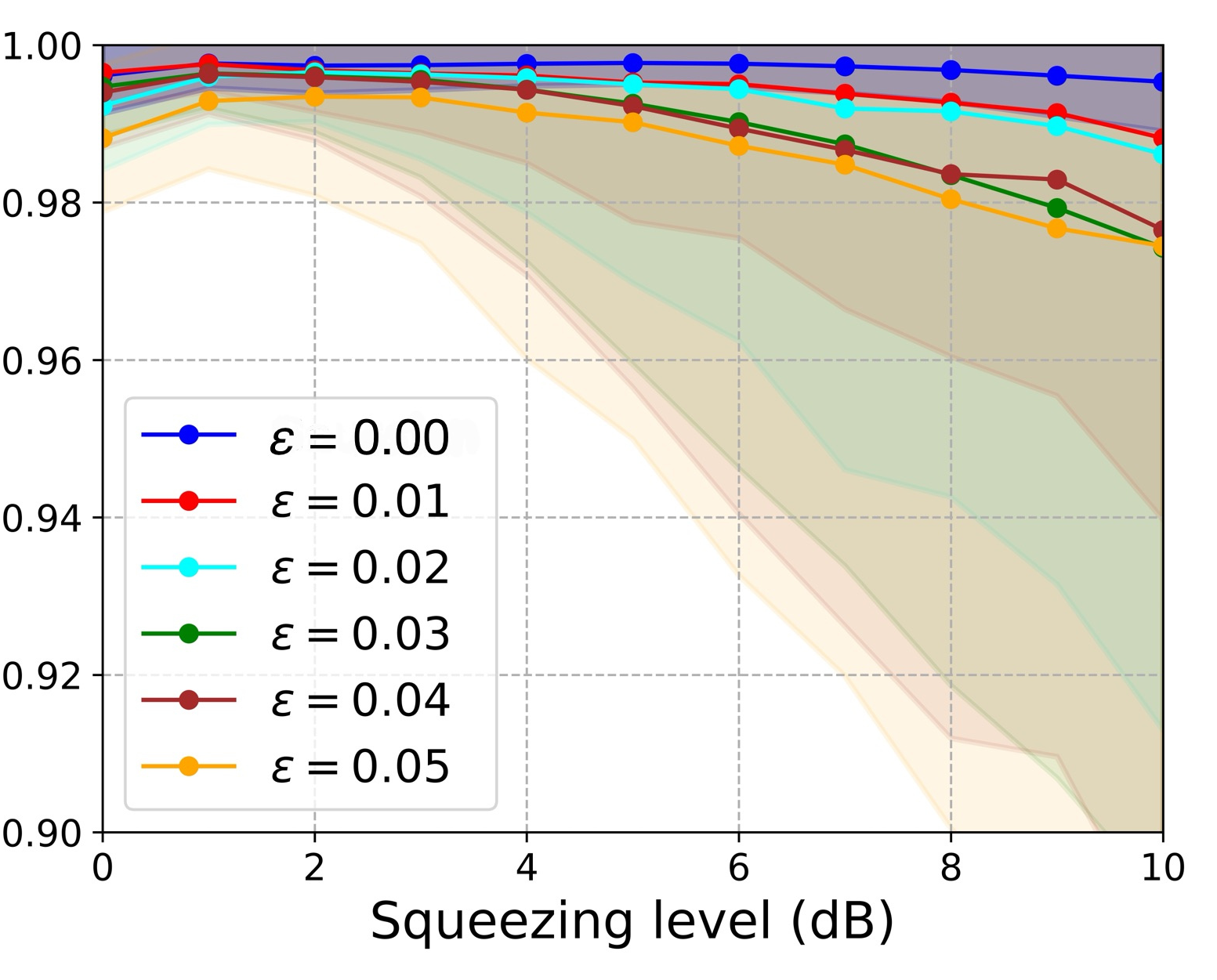}
\put(-270,142){\text{(a)}} 
\put(-90,142){\text{(b)}}  
\caption{Fidelity of the covariance-matrix-based reconstruction of single-mode squeezed states. Shaded regions in both panels indicate $\pm1$ standard deviation over the corresponding set of test states. In panel (a), our method (`$\sigma$') is benchmarked against density matrix CNN reconstruction (`$\rho$'). Panel (b) shows the average fidelity for different values of $\epsilon$. In all these cases, the fidelity remains above 0.97, which shows that, despite the presence of extra thermal effects, our method can still use the covariance matrix approach with reasonable accuracy.}
\label{Fig5}
\end{figure}

Fig.~\ref{Fig5} (a) compares the reconstruction performance of our covariance matrix estimator
($\boldsymbol\sigma$) with that of the density matrix CNN method proposed in
Ref.~\cite{Hsieh2022}. The covariance matrix estimator reaches \(\langle F\rangle = 0.99\) with a
variance below $2\times10^{-3}$, matching the accuracy of the density matrix CNN while using fewer
computational resources. Fig.~\ref{Fig5} (b) shows that fidelity remains high for all tested noise
weights: \(\langle F\rangle \ge 0.97\) at $\epsilon=0.05$ and degrades by less than 3\,\% across
the entire range. In both panels, the shaded regions represent $\pm1$ standard deviation over the
corresponding set of test states, quantifying the variability (spread) of the fidelity across
different reconstructions. This variability reflects how sensitive the reconstruction performance is
to changes in the noise parameter $\epsilon$ and the magnitude of the deviation relative to the
density matrix reconstruction. This result demonstrates the high precision of our
method, even at higher noise levels.

\section{Conclusion}

We introduced a supervised, covariance matrix–based tomography scheme for single-mode squeezed vacuum states in the presence of thermal noise, modeled as a realistic two-component mixture (squeezed-thermal plus thermal). The estimator learns a direct map from sparse quadrature sequences to a physically valid covariance matrix, without requiring full density-matrix reconstruction. Training with this experimentally motivated noise model keeps the method accurate and robust in the presence of noisy conditions inherent to the experimental environment.

On a GPU server, inferring quantum state information from a quadrature sequence takes $39\,\mathrm{ms}$. Performing 1000 estimations to construct a stable standard deviation requires only $39\,\mathrm{ms} \times 1000 =39\,\mathrm{s}$ for an experimental measurement. From the perspective of interval estimation, this is already fast. In contrast, traditional methods such as maximum-likelihood estimation (MLE) require a numerical optimization for each quantum state estimation to obtain the best density matrix, often over many iterations, making repeated bootstrapping computationally demanding. By comparison, once our covariance-matrix-based machine learning model is trained, each inference is fast, so interval estimation benefits and can be completed within an acceptable time.

Benchmarking against experiment shows that our covariance-matrix reconstructions faithfully track the SQ–ASQ degradation curve (with best agreement at $\epsilon=0.01$), capture the monotonic purity loss with increasing anti-squeezing, and yield tight $\pm1$ standard deviation error bars. The reconstructed curves follow the same trend as the power spectrum analyzer (SA) measurements. On simulated two-component states, the fidelity averages $\langle F\rangle=0.99$ with variance $<2\times10^{-3}$ and remains $\ge 0.97$ for noise weights up to $\epsilon=0.05$, demonstrating precision and robustness 
when strong thermal noise is present.

The covariance matrix offers a full characterization of lab-generated single-mode squeezed states in a sparse, compact form, in contrast to the computationally intensive density-matrix approach of Ref.~\cite{Hsieh2022}. It encodes all second-order information of a Gaussian state without requiring a Hilbert–space cutoff, thereby eliminating the dimension-dependent bias that affects Fock-basis reconstructions at high squeezing. Moreover, the key observables—squeezing (SQ), anti-squeezing (ASQ), and purity—are obtained directly from its elements with minimal post-processing. This is particularly relevant for squeezed-vacuum injection in gravitational-wave detectors, where precise measurements of quantum states are essential. In addition, the reduced model size makes the approach well suited for efficient hardware implementations such as FPGA-based real-time tomography \cite{Wu2025}.

Our analysis is based on the two-component state~$\hat{\rho}_{\text{noisy}}$ of Eq.~(\ref{Eq 6}) which helps to simulate realistic
experimental noise conditions better than using the one-component state~$\hat\rho_S$ of Eq.~(\ref{Eq 5}). 
Although noise introduces deviations from states of purely Gaussian form, the covariance matrix framework can still effectively
capture the Gaussian features embedded within such noisy states. By focusing on the 
Gaussian part of the quantum state, the covariance matrix retains the Gaussian component needed for reconstruction while 
ignoring higher-order non-Gaussian features. 

Reliable estimation of the degradation, purity, and fidelity of noisy squeezed states in one-mode systems whilst being
economical with the deployed resources, as demonstrated here, 
promises to provide further savings when extending this tomographic approach to multi-mode squeezed state systems.

\bigskip

\section*{Acknowledgments} This work is partially supported by the National Science and Technology Council of Taiwan (Nos 112-2123-M-007-001, 112-2119-M-008-007, 114-2112-M-007-044-MY3),  Office of Naval Research Global, the International Technology Center Indo-Pacific (ITC IPAC) and Army Research Office, under Contract No. FA5209-21-P-0158, and the collaborative research program of the Institute for Cosmic Ray Research (ICRR) at the University of Tokyo.

\bigskip

\section*{Disclosures} The authors declare no conflicts of interest.

\bigskip

\section*{Data availability} Data underlying the results presented in this paper
may be obtained from the author upon reasonable request.


\end{document}